\documentclass[aps,prl,twocolumn,showpacs,amsfonts]{revtex4}
\usepackage{epsfig}\usepackage{subfigure}
\usepackage{graphicx,graphics}
\usepackage{amsmath,ulem}
\usepackage{amssymb}
\usepackage{bm}
\usepackage{color}
\usepackage[utf8]{inputenc}

\newcommand{\curA}{{\cal A}}
\newcommand{\curB}{\mathcal{B}}

\newcommand{\bp}{{\bm p}}

\newcommand{\bfe}{{\bf e}}
\newcommand{\bfeb}{\bar{\bf e}}

\newcommand{\zh}{\hat{z}}
\newcommand{\sigmah}{\hat{\sigma}}

\newcommand{\hh}{\hat{h}}

\newcommand{\br}{{\bf r}}

\newcommand{\be}{\begin{equation}}
\newcommand{\ee}{\end{equation}}
\newcommand{\bea}{\begin{eqnarray}}
\newcommand{\eea}{\end{eqnarray}}
\newcommand{\bse}{\begin{subequations}}
\newcommand{\ese}{\end{subequations}}

\newcommand{\bsigma}{\bm{\sigma}}

\begin{document}

\title{Interface symmetry and spin control in topological insulator-semiconductor heterostructures}

\author{Mahmoud M. Asmar}
\affiliation{Department of Physics and Astronomy, Louisiana State University, Baton Rouge, LA 70803-4001}
\author{Daniel E. Sheehy}
\affiliation{Department of Physics and Astronomy, Louisiana State University, Baton Rouge, LA 70803-4001}

\author{Ilya Vekhter}
\affiliation{Department of Physics and Astronomy, Louisiana State University, Baton Rouge, LA 70803-4001}

\date{November 14, 2016}

\begin{abstract}

{Heterostructures combining topological and non-topological materials
constitute the next frontier in the effort to incorporate topological insulators (TIs) into
functional electronic devices. We show that the properties of the interface states appearing at the planar boundary between a
topologically-trivial semiconductor (SE) and a TI are controlled by the symmetry of the interface.
In contrast to the well-studied helical Dirac surface states, SE-TI interface states exhibit
elliptical contours of constant energy and complex spin textures with broken helicity.
We derive a general effective Hamiltonian for SE-TI junctions, and propose experimental signatures such as an out of plane spin accumulation under a transport
current and the opening of a spectral gap that depends on the direction of an applied in-plane magnetic field.}

\end{abstract}


\maketitle

{\it Introduction.} The control of topologically-protected metallic interface states~\cite{HasanKane,QiZhang,BernevigHughes} in heterostructures containing topological insulators (TIs) along with materials such as semiconductors (SE)~\cite{Zhang2012a,BerntsenPRB,Yoshimi2014}, superconductors~\cite{FU-Kane,Cava,Fou}, or magnets~\cite{Melnik,Tserk}, may allow new functionalities ranging from spintronics~\cite{Burkov2010,RaghuPRL,Pesin} to thermoelectrics~\cite{thermo2}, to quantum computing~\cite{computing} and beyond. In the analysis of the properties of such heterostructures, it is often assumed that the main features of the interface states are identical (at least in the low energy limit) to the surface states at a vacuum termination of a TI.  The
isotropic  Dirac
dispersion and helical spin-momentum locking, well established for the surface states~\cite{HasanKane,QiZhang,PanPRL2011,MiyamotoPRL2011}, are commonly ascribed to the interface states by extension.

\begin{figure} [t]
  \centering
  \includegraphics[width=.9\columnwidth]{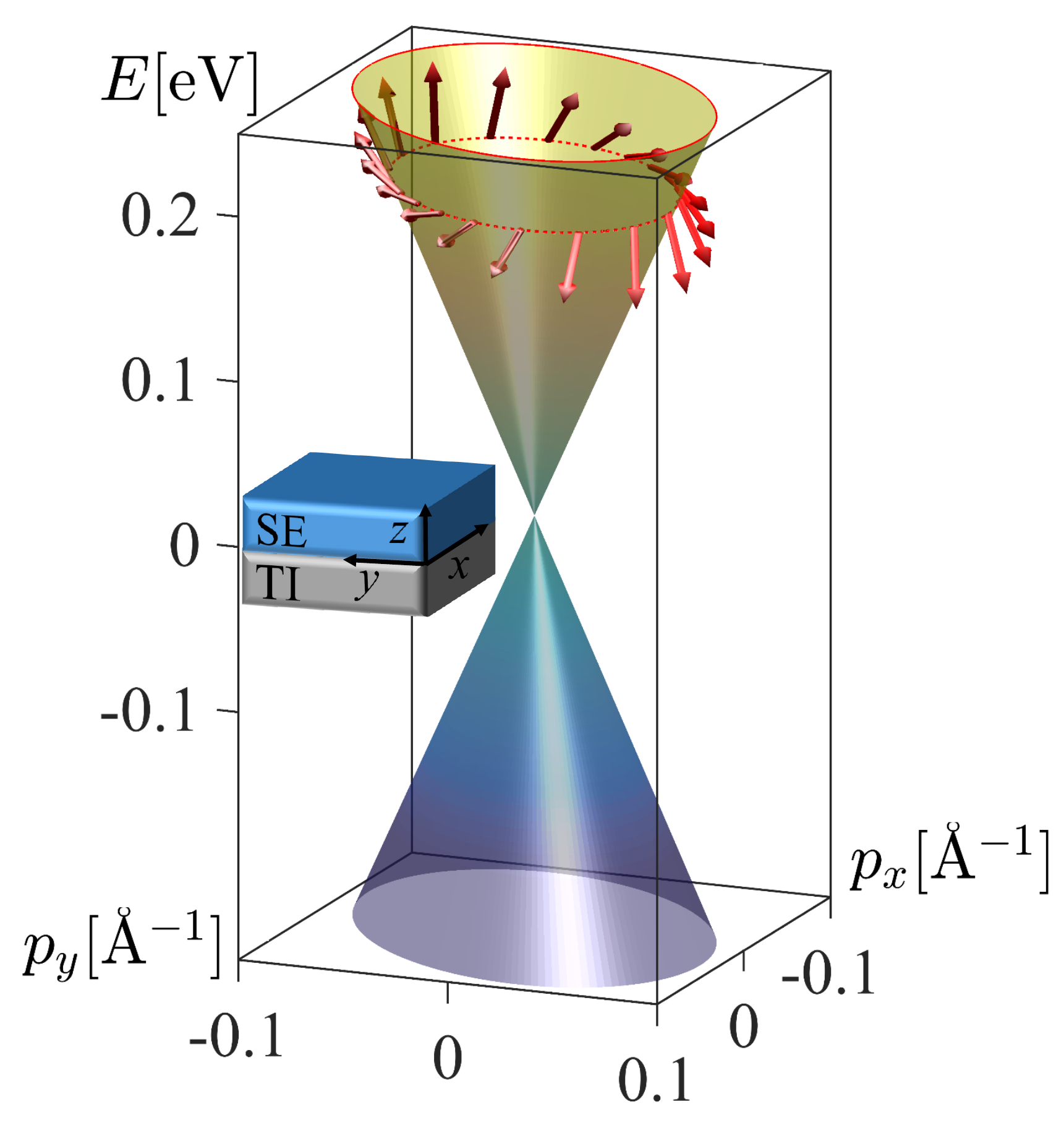}
  \caption{Generic energy dispersion and spin texture (arrows) of the topological state at a SE-TI interface. Plot is for the parameters corresponding to Bi$_2$Se$_3$, and $u_{3}=u_{4}=0.71A_1$ and $u_{5}=0.1A_{2}$ in Eq.~\eqref{eq:U}. Note the ellipticity of the constant energy contours,
  and the out of plane tilt of the spins. Inset:  Geometry of the SE-TI junction.
}
    \vspace{-0.75cm}
    \label{fig1}
\end{figure}

We show that the these features
of the
TI surface states require protection by
spatial symmetries that can be broken at interfaces.
We consider
allowed nonmagnetic interface potentials that can break such spatial symmetries (while preserving time-reversal (TR) symmetry), and
derive the most general
low-energy Hamiltonian for SE-TI interfaces. We
find that
the interface states
exhibit broken helicity, with spins rotating out of the plane of the interface,
and broken in-plane rotational symmetry leading to an
elliptical Dirac-like energy spectrum, as summarized in Fig.~\ref{fig1}.
As a result, an in-plane Zeeman field may not only shift the Dirac cone~\cite{BurkovBfield}, but also open a field-orientation-dependent gap in the spectrum. In addition, under an in-plane transport current, the spin accumulation often has an out-of-plane component,
broadening the range of potential applications.

\begin{table}[t] 
  \centering
\begin{tabular}{||c|c c c c| c |c ||}
  \hline \hline
    Irrep &E& C$_2$&$\sigmah_{\bfe}$&$\sigmah_{\bar{\bfe}}$ &Term in $H_I$ & Term in $\widehat{U}$\\ [0.5ex]
  \hline\hline
 $\curA_1$  &1 &1  & 1 & 1  &  $\Delta,v_1,v_4$ & $u_0,u_1,u_2$  \\
 $\curA_2$  &1 &1  & -1 & -1  &   $v_2,v_6$ &$u_5$  \\
  $\curB_1$&1 &-1  & 1 & -1  & $v_5$ & None \\
 $\curB_2$ &1 &-1  & -1 & 1  & $v_3$&$u_{3},u_4$    \\
  \hline\hline
\end{tabular}
  \caption{Character table of the irreducible representations  of C$_{2v}$.
  Here, E is the identity operation, C$_2$ is two-fold rotation, and $\sigmah_\bfe$ and $\sigmah_{\bfeb}$ are
mirror reflections with respect to the $\bfe$-$\zh$ and $\bfeb$-$\zh$ planes, respectively.
  The third and  fourth columns classify the symmetries of the interface Hamiltonian, Eq.~(\ref{hvsPP}), and the interface potential $\widehat{U}$, Eq.~(\ref{eq:U}), respectively.}
\label{charinterface}
\end{table}

{ \it Symmetries of a SE-TI junction.}
We consider a flat planar interface, $z=0$, between a semi-infinite three-dimensional TI and a topologically-trivial SE,
see Fig.~\ref{fig1} (inset), that supports two-dimensional
states~\cite{Zhang2009,MooreBalents,Liu2010}.
In the well-studied TI-vacuum case the interface states are described by a
Dirac Hamiltonian in spin space,
\be
\label{Eq:hdirac}
H_{D} = v(\bsigma\times\bp)_z\,,
\ee
where $\bp=(p_x, p_y)$ is the in-plane momentum,  $v$ is the
effective velocity ($\hbar = 1$)  and $\bsigma$
is a vector of the Pauli matrices.
The eigenstates of $H_D$ have a linear dispersion, $E(\bp)=\pm v|\bp|$, with circular constant energy contours, obeying full rotational symmetry around the $\bm \zh$ axis, C$_\infty$. In addition, $H_D$ commutes with with the helicity operator, $\hh = (\bsigma \times\bp)_z /p$, and hence the eigenstates are helical,
with the spin expectation value
in the plane and perpendicular to ${\bm p}$.
These are the symmetries of
the low energy limit of the  $\bm k\cdot \bp$ Hamiltonian for bulk TI materials, such as Bi$_2$Se$_3$ or Sb$_2$Te$_3$~\cite{Zhang2009}.  Thus, Eq.~(\ref{Eq:hdirac}) implicitly assumes that no additional symmetries are broken by the interface~\cite{Isaev,Enaldiev2015}.

Crucially, a SE-TI interface is likely to have lower symmetry than
the bulk materials on each side,~\cite{Voisin2,Zunger}
strongly modifying electronic properties.
The most general linear in $\bp$ Hamiltonian that is invariant under TR ($\bsigma\to - \bsigma$
and $\bp \to - \bp$),
\be
H_{I} =  \sigma_0 \Delta +  \sum_{m=x,y,z}\sum_{n=x,y} c_{mn}\sigma_m p_n\,,
\label{eq:genericsurface}
\ee
is characterized by seven real parameters: An overall energy shift $\Delta$, and the coefficients $c_{mn}$.
Here $\sigma_0$ is
the $2\times 2$ identity matrix. Note that, while the momenta are constrained to the plane of the interface, electron spins can have a $z$-component.

The Hamiltonian, Eq.~(\ref{eq:genericsurface}), has a much lower symmetry than $H_D$.
Continuous rotational symmetry around the $\bm \zh$ axis, $C_\infty$, is conspicuously absent.  The helicity is also broken. However, $H_I$ is particle-hole symmetric
(relative to the energy shift $\Delta$) since the anticommutator $\{\bsigma\cdot \bm m, H_I-\sigma_0\Delta\}=0$  for the vector $\bm m$ with $m_x=c_{yx} c_{zy}-c_{yy}c_{zx}$, $m_y=c_{xy} c_{zx}-c_{xx}c_{zy}$, and $m_z=c_{xx} c_{yy}-c_{xy}c_{yx}$.
Moreover, the spins of all the eigenstates of $H_I$ are in the plane normal to $\bm m$~\cite{SM}.

In typical models of a TI-vacuum interface, $\bm m\|\widehat{\bm z}$, maintaining
the $C_\infty$ symmetry.
When $\bm m$ points away from the $z$-axis, the group of rotations is reduced. The remaining symmetry depends on the spin-orbit structure of the interface potential.

If $\bf e$ is a vector in the plane of the interface (given its physical meaning below by the interface potentials), we define $\bar{\bf e}=\widehat{\bm z}\times {\bf e}$ and rewrite Eq.~\eqref{eq:genericsurface} as~\cite{SM}
\bea
\label{hvsPP}
&&H_I =\Delta\sigma_0+  v_1 (\bsigma \times \bp)_z + v_2 \bsigma \cdot \bp + v_3 \sigma_z\bp\cdot \bfe
\\
&&\quad
+ v_4
(\bsigma\cdot \bar{\bf e})(\bp\cdot \bfe)
 + v_5  \sigma_z (\bp\cdot \bar{\bf e})
+ v_6 (\bsigma\cdot \bfe) (\bp\cdot \bfe)\,.
\nonumber
\eea
The advantage of this form is that each term in Eq.~(\ref{hvsPP}) belongs to a particular irreducible representation (irrep) of the C$_{2v}$ group, as listed
in Table~\ref{charinterface}. It is clear that the spin structure of the eigenstates of $H_I$ is complex. For example, the $v_1$ and $v_2$ terms favor spin orientation normal to and along $\bm p$ respectively, while the $v_3$ and $v_5$ terms tend to tilt the spins out of the plane of the interface.
At low energies the only source of symmetry breaking is the interface potential, and Eq.~\eqref{hvsPP} gives the most general Hamiltonian for the analysis of the interface states. Hence, controlling the parameters $v_i$ allows control of the spin properties of the interface state, and we now compute those within a specific model.

{\it Model of a SE-TI junction.}
We model the interface in Fig.~\ref{fig1} by the Hamiltonian
\be
\label{Eq:hamiltonian}
H =  H_{\rm TI}\Theta(-z)  +H_{\rm SE}\Theta(z)  + \widehat{U}\delta(z)\,.
\ee
Here $H_{\rm TI}$ ($H_{\rm SE}$) describes a bulk topological insulator (semiconductor) at $z<0$ ($z>0$), and
$\Theta(z)$ is the Heaviside step function.
The last term
describes the interface potential,  %
and the choice of the delta-function simplifies calculations without loss of generality.

We take $H_I$ to be the model Hamiltonian for Bi$_2$Se$_3$~\cite{Liu2010}, written as a $4\times 4$ matrix in the basis of
column vectors
$\psi = (\psi_{+\uparrow}, \psi_{-\uparrow},  \psi_{+\downarrow},  \psi_{-\downarrow})^{T}$,
where the subscripts $\pm$ ($\uparrow,\downarrow$)  refer to parity (spin),
\begin{eqnarray}
\nonumber
 H_{\rm TI}&=&
\sigma_0\otimes[\tau_{z}(M-B_{1}p_{z}^2-B_{2}p^{2})+A_{1}\tau_{y}p_{z}]
\\
&& \qquad +A_{2}
(\bsigma\times \bp)_{z}\otimes\tau_{x}\,.
\label{Bhz3d}
\end{eqnarray}
Here $A_i,B_i$, ($i=1,2$) are material-dependent
constants, $M>0$ determines the TI
bulk band gap, $\tau_i$
is the Pauli matrix in
the parity space, and $\otimes$ denotes a direct product.
Since it is the relative sign
of $M$ and $B_i$ in Eq.~\eqref{Bhz3d} that determines the topological
nature of the insulator, we take $H_{\rm SE}$ to differ from $H_{TI}$ only by the sign of
 mass parameter ($M\to -m<0$).
All the results are obtained using the parameters for Bi$_{2}$Se$_{3}$  from Ref.~\onlinecite{Liu2010}: $A_{1}=2.26\; eV$\AA , $A_{2}=3.33\; eV$\AA
, $B_{1}=-6.86\; eV$\AA$^{2}$, $B_{2}=-44.5\; eV$\AA$^{2}$,
and $M=-m=0.28\ eV$.
Note that both $H_{\rm TI}$ and $H_{\rm SE}$ have particle-hole symmetry since, for ${\cal P}=\sigma_z\otimes\tau_x$, we have $\{{\cal P},H_{\rm TI/SE} \}=0$.

The symmetry of the matrix $\widehat{U}$
controls
the nature of the low-energy interface states.
This matrix has 16
independent coefficients corresponding to $\sigma_i\otimes\tau_j$ with $i,j=0,x,y,z$. For a non-magnetic interface the requirement that
the commutator $[\widehat{U}, \widehat{T}]=0$,
where the time-reversal operator, $\widehat{T}=i\sigma_{y}\otimes\tau_0\mathcal{C}$,
and $\mathcal{C}$ denotes complex conjugation,
leaves only six allowed independent terms
\be
 \widehat{U}=\!\sigma_0\otimes\left[u_0\tau_0\!+\!u_1\tau_z\!+\!u_2\tau_x\right]\!+\!
 \left[u_3\sigma_x\!+\!u_4\sigma_y+u_5\sigma_z\right]\otimes\tau_y\,.
\label{eq:U}
\ee
Here, $u_0\pm u_1$ denotes
potential scattering for even/odd parity states,
while $u_2$
describes parity mixing due to the interface
potential (note that parity is explicitly broken at the SE-TI
interface).  Due to spin-orbit coupling, lowering the in-plane spin rotational symmetry breaks real-space rotations as well.
Consequently, the $u_3$ and $u_4$ potentials
define the vector $\bfe$ according to
\be
\label{eq:qdef}
u_3\sigma_x+u_4\sigma_y = q \bfe\cdot \bsigma,\qquad q = \sqrt{u_3^2+ u_4^2}\,.
\ee
We see below that this is precisely the vector introduced in Eq.~\eqref{hvsPP}. Any potential that modifies  the spin-orbit term at the interface (strain, interdiffusion of atoms, etc.) yields such terms.  Finally, $u_5$ is the spin-orbit term that breaks mirror symmetries but preserves rotations.

Notably, as seen in Table~\ref{charinterface}, the various terms in
$\widehat{U}$ can also be classified according to the C$_{2v}$ point group, showing
the connection between interface potentials and terms in $H_I$. We now derive Eq.~(\ref{hvsPP}) from
the
Hamiltonian of a SE-TI interface, Eq.~\eqref{Eq:hamiltonian}.

{\it Effective Interface Hamiltonian.}
Using Eq.~\eqref{Eq:hamiltonian} we determine the energies, $E_i$, and normalized eigenfunctions, $|\Psi_i\rangle$, of the states localized at the interface, and form the
matrix representation of the interface Hamiltonian $H_I=\sum_i E_i|\Psi_i\rangle\langle\Psi_i|$.
We choose the states of
of $H_{\rm TI}$ and $H_{\rm SE}$
in the helical representation~\cite{Isaev},
\begin{equation}\label{states}
\psi_{t\nu\mu}(\br,z)= \begin{pmatrix}
                                     ia_{t\nu\mu} \\
                                     ib_{t\nu\mu} \\
                                     t a_{t\nu\mu}e^{i\theta_{\bp}} \\
                                    t b_{t\nu\mu}e^{i\theta_{\bp}}  \\
                                   \end{pmatrix}e^{i\bm p \cdot \bm r} e^{\lambda_{\nu\mu}z}\,.
\end{equation}
Here $t = \pm 1$ is the helicity quantum number, ${\bf r}$ is the in-plane coordinate,
$\tan\theta_{\bp}=p_y/p_x$,
$\mu=B$ ($\mu=T$) denotes the TI, Bottom (SE, Top) side of the interface,
\begin{subequations} \label{as}
\begin{eqnarray}
  a_{t\nu\mu} &=& A_{1}\lambda_{\nu\mu}-t A_{2}p \;,\\
  b_{t\nu\mu} &=& M_{\mu}+B_{1}\lambda^{2}_{\nu\mu}-B_{2}p^{2}-E\,,
\end{eqnarray}
\end{subequations}
$M_T = M$, and $M_B = -m$. For each $E,\bp$ there are two allowed values of the decay exponent  $\lambda_{\nu\mu} (E,\bm p)$
(labelled by $\nu=\pm$) on each side of the interface, which are the roots of
\be
\label{Eq:eigenvaluelambda}
E^2-\mathcal{M}_{+\mu}\mathcal{M}_{-\mu}-A_{2}^2p^2=0\,,
\ee
where $\mathcal{M}_{\nu\mu}=M_{\mu}+B_{1}\lambda^{2}_{\nu\mu}-B_{2}p^{2}\pm A_{1}\lambda_{\nu\mu}$.
For interface states $Re[\lambda_{\nu T}]<0$ and $Re[\lambda_{\nu B}]>0$.

\begin{figure} [t]
    \begin{center}
        \subfigure{
            \includegraphics[width=45mm,height=45mm ]{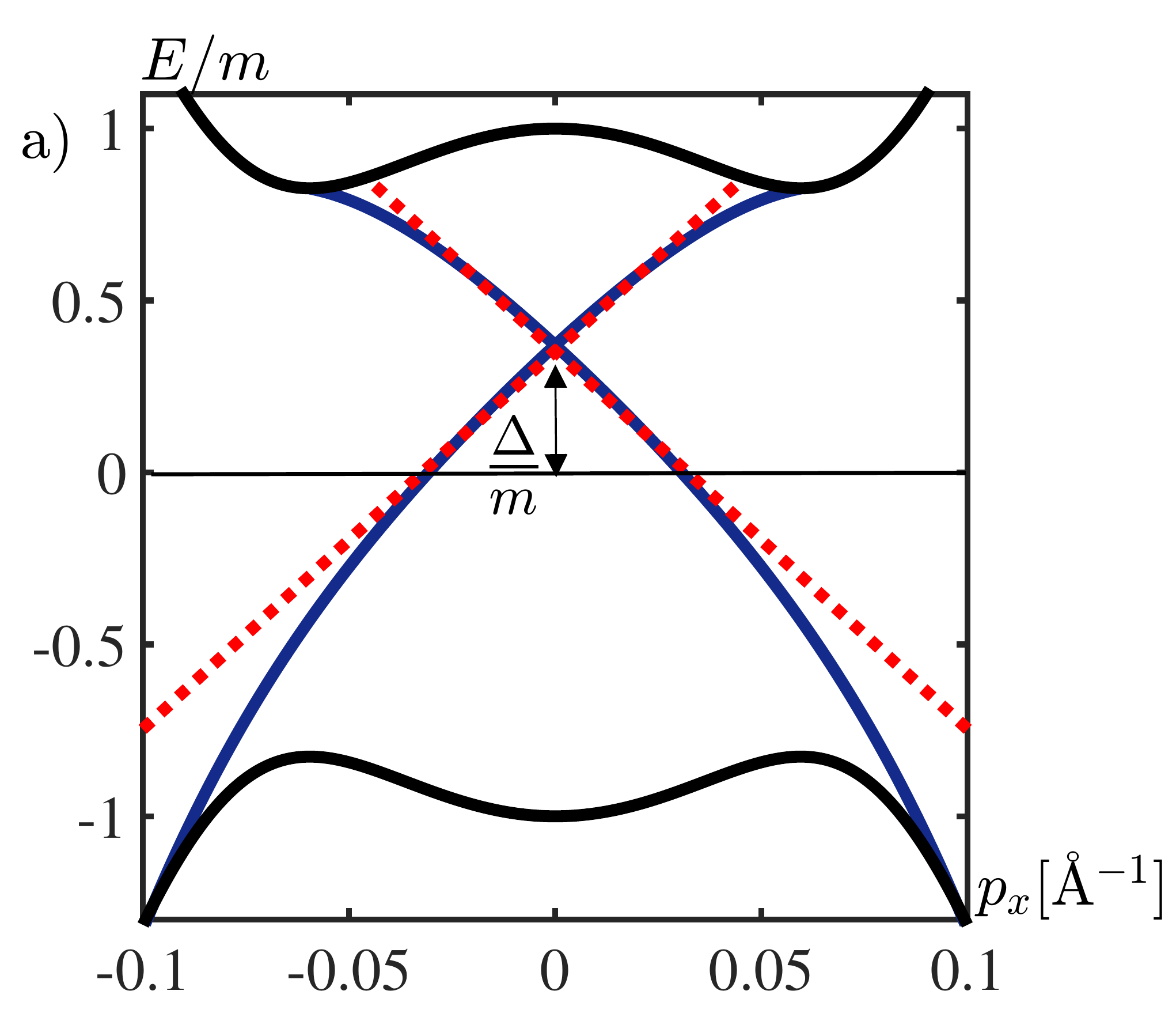}}%
        \subfigure{
            \includegraphics[width=45mm,  height=45mm ]{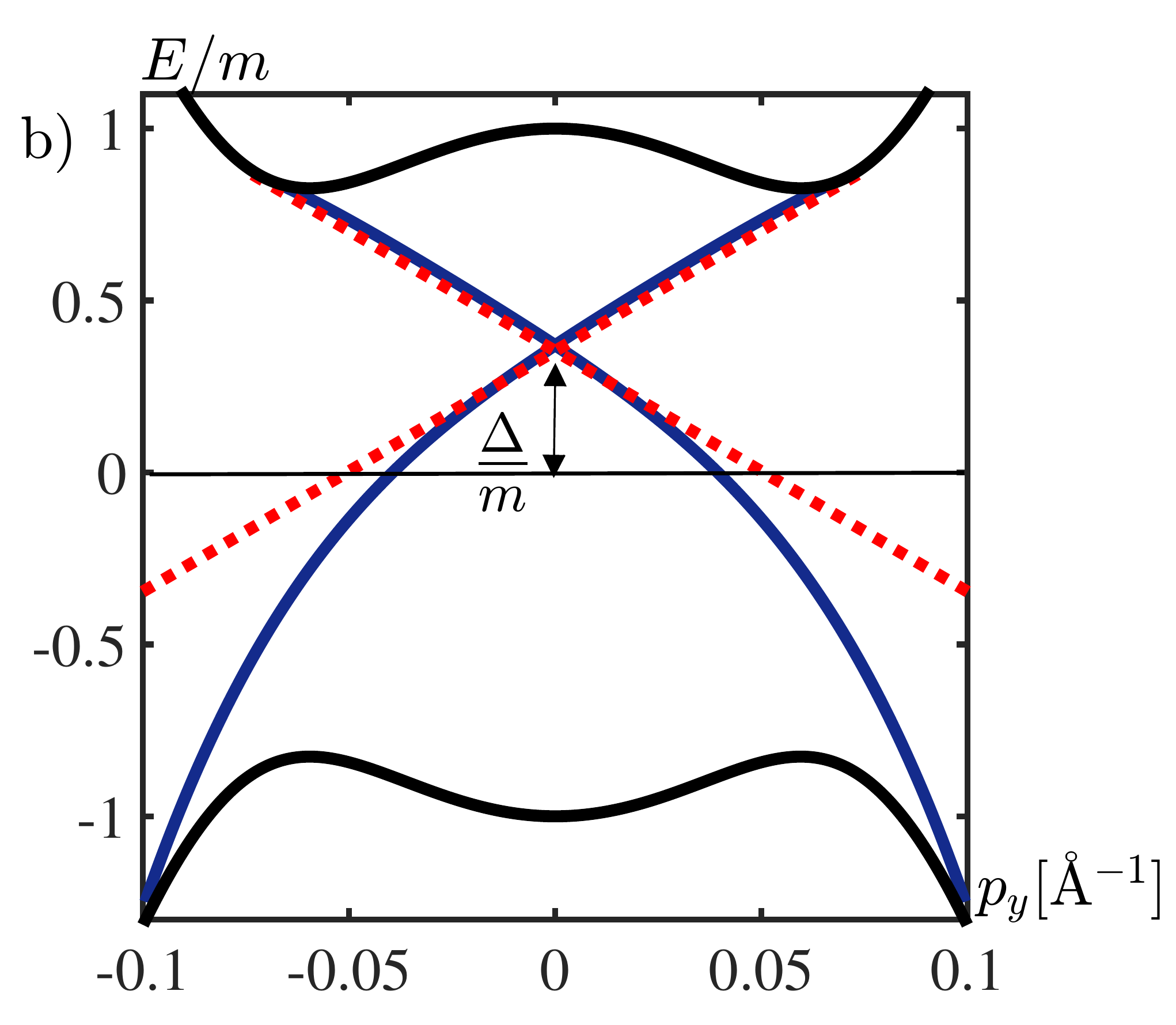}}
                                \subfigure{
            \includegraphics[width=45mm, height=45mm]{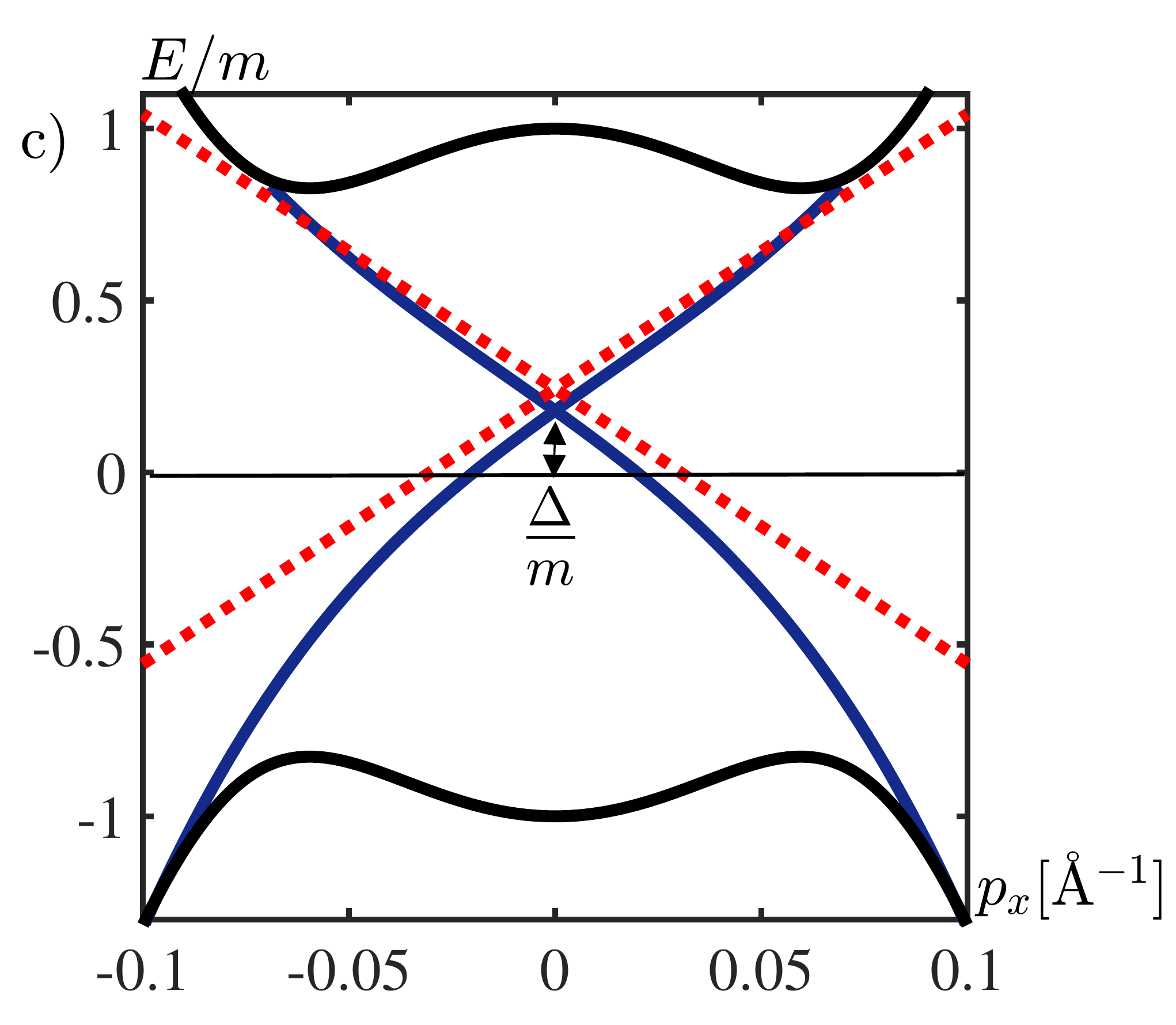}}%
                    \subfigure{
            \includegraphics[width=45mm, height=45mm]{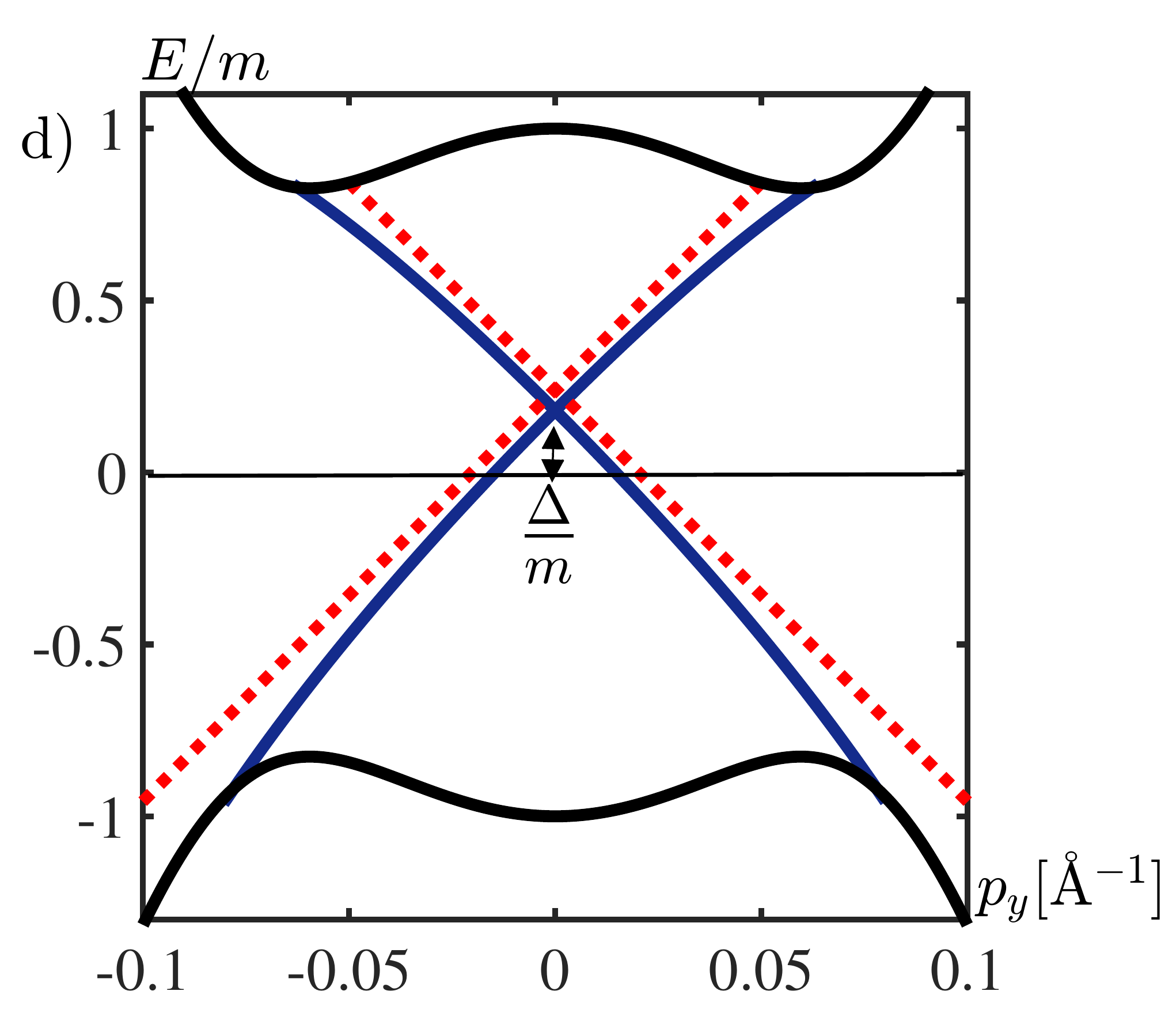}}
    \end{center}
    \vspace{-.5cm}
    \caption{
Numerical (solid lines) and approximate analytical (dashed lines) dispersion curves along the $p_y=0$ (left column, panels (a) and (c)), and $p_x=0$ (right column, panels (b) and (d)) directions. The choice of interface potentials is $u_{0}=0.6A_{2}$, $u_{2}=-0.3A_2$, $u_{1}=0.1A_2$, $u_{5}=0.15A_2$, $q=0.9A_2$ and $\phi=0$ for panels (a),(b), and
$u_2=0.3A_{2}$, $q=0.3A_2$ and $\phi=\pi/2$, with all other $u_{i}=0$ for panels (c),(d).} %
\label{fig:shifts}
\end{figure}

The eigenfunctions of
Eq.~\eqref{Eq:hamiltonian}, $H\Psi(\br,z) = E\Psi(\br,z)$, at each energy $E$
and momentum $\bp$, can be written as $\Psi(\br, z)=\Psi_B(\br, z)\Theta(-z)+\Psi_T(\br, z)\Theta(z)$ with
\be
\label{eq:psienergy}
\Psi_{\mu}(\br, z) = \sum_{t=\pm}\sum_{\nu=\pm}C_{t\nu\mu}\psi_{t\nu\mu}(\br,z)\,.
\ee
The coefficients $C_{t\nu\mu}$ in Eq.~\eqref{eq:psienergy} as well as the dispersion, $E(\bp)$ of the interface states are determined from the boundary conditions. The continuity of the wave function requires
$\Psi_B(\br,0) = \Psi_T(\br,0)\equiv\Psi(\br,0)$, while the interface enforces the discontinuity in the derivatives,
\be
\label{derivative}
 B_{1}\sigma_0\otimes\tau_{z}\big[\partial_{z}\Psi_{B}(\br,0)-\partial_{z}\Psi_{T}(\br,0)\big]=\widehat{U}\Psi(\br,0)\,.
\ee
Since $\psi_{t\nu\mu}$ are four-component functions,
these boundary conditions take the form of $\mathcal{B} \bm x=0$ where $\mathcal{B}$ is an $8\times 8$ matrix and $\bm x$ is a
vector of the $C_{t\nu\mu}$. The roots of $\det \mathcal{B}=0$ give $E(\bp)$, while the eigenvectors $\bm x_E (\bp)$ determine the eigenfunctions, Eq.~\eqref{eq:psienergy}.
We note that neglecting the second derivative,   $p_z^2 = -\partial_z^2$, in Eq.~\eqref{Eq:hamiltonian} and allowing for a discontinuity in $\Psi(\br,z)$
at the topological boundary~\cite{Zhang2012} gives results qualitatively different from ours.

\begin{figure} [t]
  \centering
  \includegraphics[width=\columnwidth]{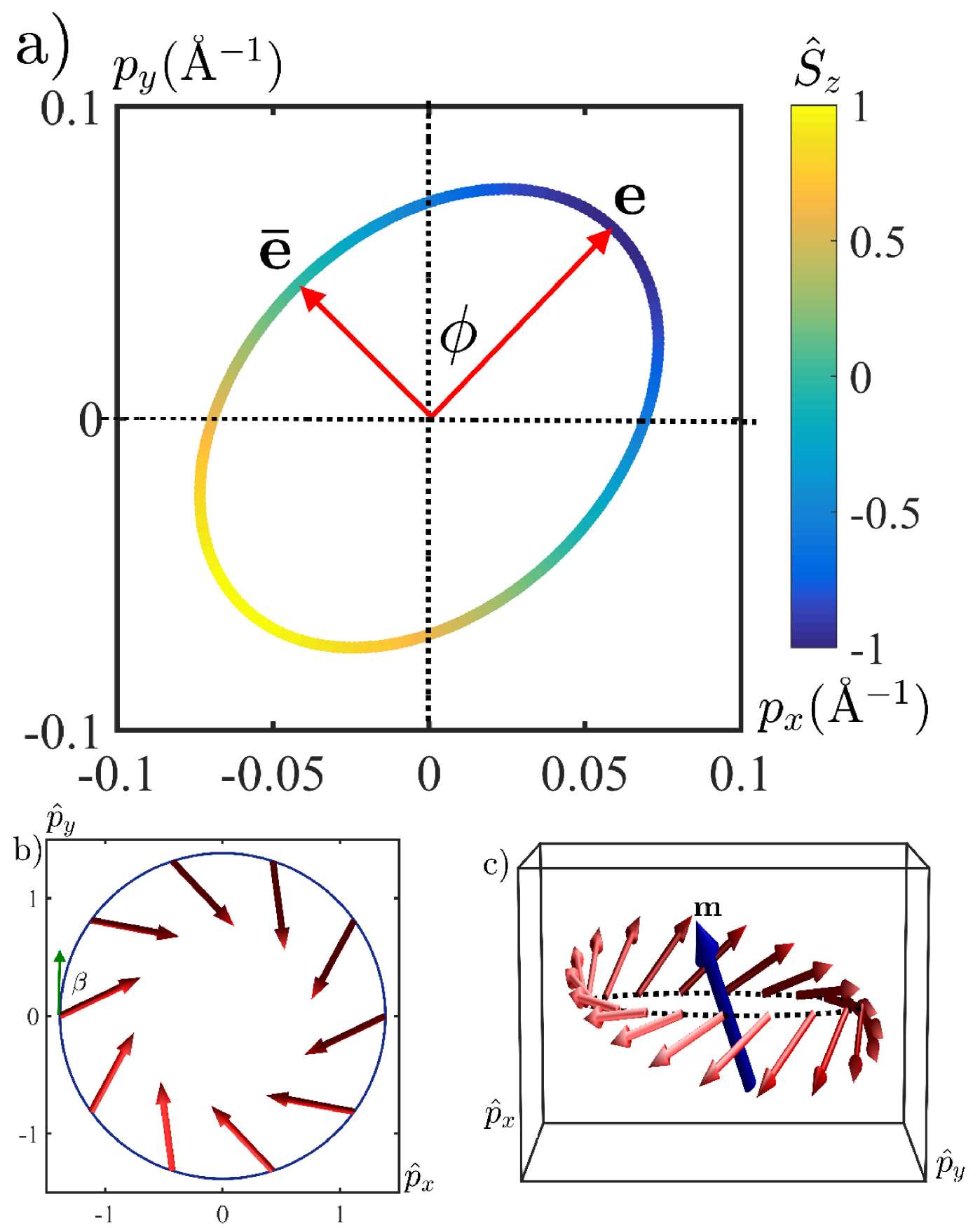}
  \caption{Constant energy surface and spin texture of the interface states.
Panel (a) shows the remaining twofold symmetry with the vectors $\bfe$ and $\bfeb$ along the semi-axes of the elllipse, see text. Color indicates the $z$ component of the spin, $\hbar\hat{S}_z/2=\langle\sigma_z\rangle$ with
  $E=2m$eV and
  other  parameters the same as in Fig.\ref{fig1}.
Panel (b): effect of broken helical symmetry due to the $\curA_2$ irrep potential
$u_5=A_2$. Panel (c):  All the spins for $E=2 m$eV, for $q=0.45A_{2}$ and $\phi=\pi/4$ ($u_{3}=u_{4}=0.32A_{2}$)
are normal to a vector $\bm{m}$ described in text. Here $\widehat{p}=A_2p/E$.}
   \vspace{-0.5cm} \label{fig2}
\end{figure}

To derive an analytic form of the interface Hamiltonian valid at low energies,we
expand $\lambda_{\nu\mu}(E, \bm p)$ around its value at $E=0,\bm p=0$, and evaluate  $\det \mathcal{B}$ to second order in $E$ and $\bp$. The resulting lengthy expressions
directly give the the coefficients $\Delta$ and $v_i$ of Eq.~(\ref{hvsPP}) in terms of the parameters of the bulk Hamiltonian and the interface potential $\widehat{U}$~\cite{SM}.
Fig.~\ref{fig:shifts} shows excellent agreement of the dispersion obtained in this linearized approximation
with the numerical solution, and clearly demonstrates broken rotational symmetry of the spectrum.

This approach makes clear the reduction from the $4\times 4$ band basis to the $2\times 2$ form of $H_I$. Only two of the
components of the eigenstates $\Psi$ are linearly independent: those correspond to different spin eigenstates (protected by the TR invariance), but mix the parities of the wave functions. As a result, the Hamiltonian has only four non-vanishing components and takes the form of Eq.~(\ref{hvsPP}). According to Table~\ref{charinterface} no part of the interface potential in our model belongs to the $\curB_1$ irrep,
and hence $v_5=0$.

{\it Representative results.}
Since the interface is characterized by six coefficients, $u_i$ in Eq.~\eqref{eq:U}, there is a large parameter space to explore, and here we
focus on
conceptually important cases.
First, for any $u_i\rightarrow\infty$, we recover the pure Dirac Hamiltonian since strong interface potential decouples the two sides of the heterostructure, reducing the problem to that of a TI surface.

If only $u_0$, $u_1$, and $u_2$ are non-zero,
the interface potential is spin-independent and belongs to the $\curA_1$ representation.
For our choice of $\widehat U$
all the terms in $\curA_1$ preserve the helicity, while
the $v_4$ term in  Eq.~\eqref{hvsPP} does not, and hence must vanish~\cite{v4}. Consequently the interface Hamiltonian again reduces to the Dirac form
\be
H_I =\Delta\sigma_0+  v_1 (\bsigma \times \bp)_z\,.
\ee
Since the $u_1$ term anticommutes with ${\cal P}$, the shift,
$\Delta$,  is induced only by $u_0$ and $u_2$~\cite{SM}.

Perfect helicity is broken for time-reversal-invariant, but spin-dependent, potentials.
Consider  $\widehat U=u_5\sigma_z\otimes\tau_y$, with all other $u_i=0$, so that it
is in the $\curA_2$ irrep. Since $\{{\cal P}, H\}=0$, there is no shift of the Dirac cone, $\Delta=0$. In addition,  $C_\infty$ is preserved, and hence $v_4=v_6=0$. The  remaining terms in the $\curA_1$ and $\curA_2$ representations give
\be
H_I = v_1 (\bsigma \times \bp)_z + v_2 \bsigma \cdot \bp  =
v'(\bsigma \times \bp)_z{\rm e}^{i\sigma_z\beta}\,.
\ee
The competition between the two terms results in the Dirac Hamiltonian with the renormalized velocity,
$v'=\sqrt{v_1^2+v_2^2}$, and
an additional global
spin rotation by an
angle $\beta = \tan^{-1}(v_2/v_1)$ around the $z$-axis. The spins of the eigenstates stay in the plane, but are no longer normal to the momenta,  as depicted in Fig.~\ref{fig2}(b). The coefficient $v_2$, and the deviation from the helicity, $\tan\beta$, are both proportional to $u_5$.

The breaking of rotational symmetry requires $u_{3,4}\neq 0$, and we now consider
this case, with all other $u_i=0$.
Then the $C_2$ and one of the mirror symmetries are broken, see Table~\ref{charinterface}, implying
\be
\label{eq:hu3u4}
H_I = v_1 (\bsigma \times \bp)_z  + v_3 \sigma_z
\bp\cdot \bfe
+ v_4 (\bsigma\cdot \bar{\bf e})(\bp\cdot \bfe),
\ee
where $v_1$, $v_3$ and $v_4$ depend on $u_i$~\cite{SM}.  The powers of $\bf e$ in this expression correspond to the powers of $q$ from Eq.~\eqref{eq:qdef}, so that $v_3\propto q$, while $v_4\propto q^2$. The former induces a spin rotation
around the $\bfe$ axis (as
illustrated in Fig.~\ref{fig2}(c)), which leads to tilting of spins out of the plane of the interface.  At the same time
the constant energy contours stretch along the direction of ${\bf e}$
and become elliptical, as
shown in Fig.~\ref{fig2}(a).

 Simple power counting arguments (confirmed by our calculations) show  that $v_6\propto u_5 q^2$ only appears in the presence of two different symmetry-breaking components  of $\widehat{U}$ (in the $\curA_2$ and $\curB_2$ irreps). The expressions for $v_i$ become complex when all $u_i\neq 0$~\cite{SM}, but
the same conclusions
remain valid.   The essential advantage of the form of Eq.~\eqref{hvsPP} over Eq.~\eqref{eq:genericsurface} is precisely the ability to predict the functional dependence of the coefficients of $H_I$ from the
symmetry of the interface.

{\it Experimental Consequences.}
The key qualitatively different predictions of our analysis include the rotated helical spin texture in the presence of
the $\curA_2$ potentials, and the elliptical shape of the constant energy contours and out-of plane spin orientation
for $\curB_2$ potentials.
These yield unexpected and non-trivial experimental consequences.

First, for perfectly helical states, the application of a magnetic field parallel to the interface shifts the Dirac cone in  momentum space, but does not change the spectrum or the helicity \cite{BurkovBfield}. In contrast, if the spin of the interface state has an out-of plane component, as for the
SE-TI interfaces with the $\curB_2$ interfaces,
Eq.~(\ref{eq:qdef}), an
in-plane field, $ { \bf B}$, opens a spectral  gap, $\delta({ \bf B})$.

\begin{figure} [t]
  \centering
  \includegraphics[width=\columnwidth]{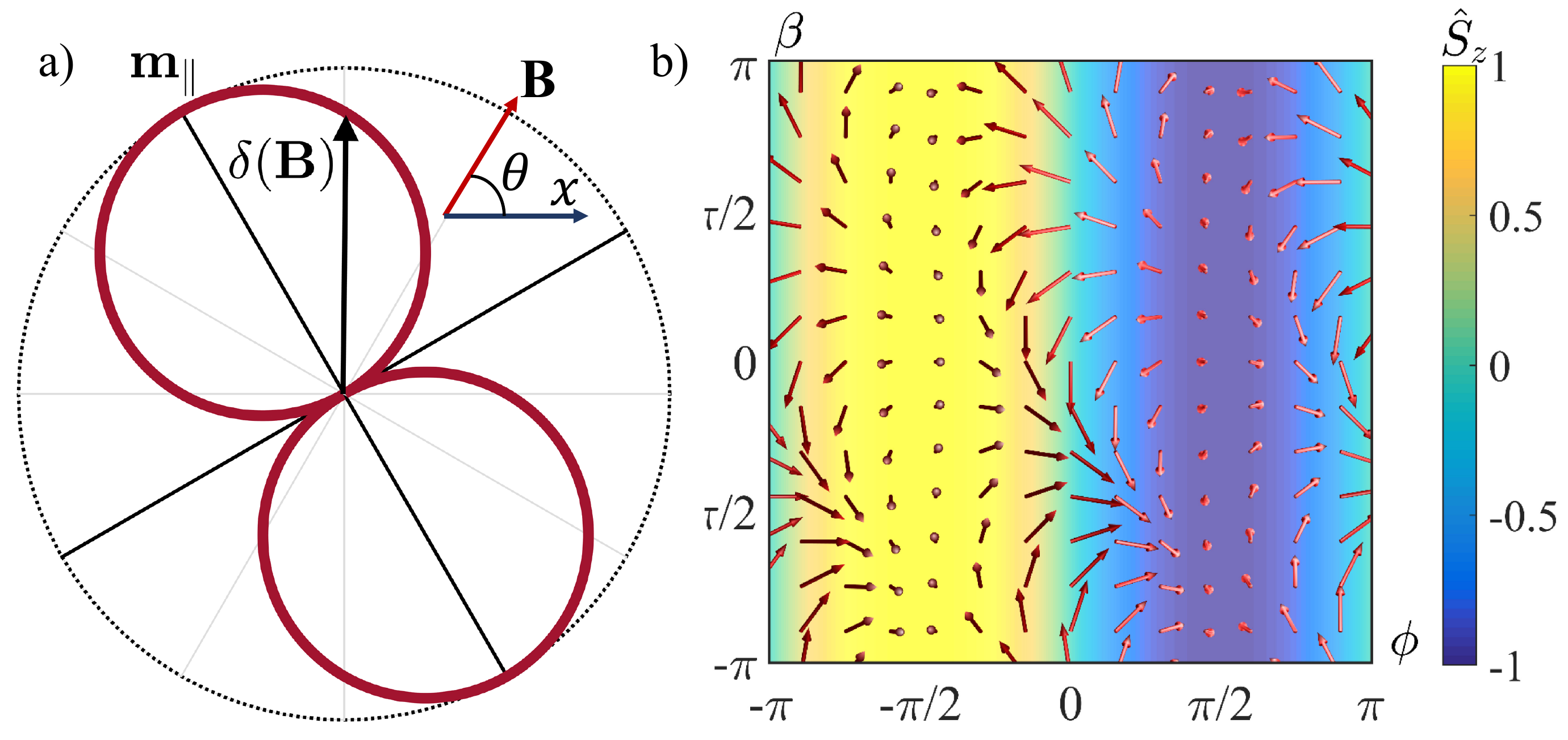}
  \caption{Experimental consequences of the broken helicity and anisotropy of the Dirac cone. a) Polar plot of the gap generated by an in plane magnetic field as a function of the field direction. $\bm m_\|$ is the in-plane projection of $\bm m$.
  b) Spin accumulation under transport current along the $x$-axis as a function of the in-plane helicity deviation angle, $\beta$, and the angle between the major semi-axis of the elliptical dispersion, $\bf e$, and the $y$-axis, $\phi$.
 Arrows and color indicate the  spin orientation, and the out of plane component of spin respectively. The full $z$-axis polarization is achieved for current along ${\bf e}$, while for the current along $\bar{\bf e}$ the magnetic moment is entirely in the plane. For panel b) $q=0.9A_{2}$.  }
  \label{figFour}
\end{figure}

If ${ \bf B}$ is at an angle $\theta$ to the $x$-axis,  $\delta({ \bf B})=2 |\bm h\cdot \bm{\widehat m}|$, where $\bm h=g\mu_B {\bf B}$, $\bm{\widehat m}=\bm m/|\bm m|$, and  $\bm m$ is the vector
normal to the spins at the interface, given above.
The gap is maximal for the field along (or opposite to) the in-plane projection of $\bm m$, and vanishes
when ${\bf B}$ is normal to that direction, see Fig.~\ref{figFour}a). This result is consistent with a
gapless spectrum for $\bm m\|\widehat{\bm z}$.
Detection of this orientation-dependent field-induced gap (shown in Fig.~\ref{figFour}a))
tests for
the symmetry breaking $u_{3,4}$ potentials.

Second, as seen from Eq.~\eqref{hvsPP},
the components $j_i$ of the current operator
are proportional to the spin density~\cite{Pesin} along a direction that depends on the coefficients of $H_I$.
Therefore, a transport current generates net spin magnetic moment.

For the usual helical surface states
the  direction of this moment is in the plane and normal to  the current~\cite{Burkov2010,RaghuPRL}. Such an accumulation was
observed experimentally~\cite{LiNatNano,Dankert,TianSciRep}.
We find that
the spin structure depends on
the interface.
As discussed above,
the $u_5$ potential leads to an in-plane spin rotation (hence the net magnetic moment acquires a component along the current), while the
$u_3$ and $u_4$ potentials rotate the spins out of the interface plane, leading to an accumulation of the $z$-component of the spin. This behavior is shown
in Fig.~\ref{figFour}b).
Putting the transport current along the
vector $\bf e$, see Fig.~\ref{fig2} yields a complete polarization normal to the interface. Setting a current along the $\overline{\bf e}$ direction, in contrast, yields an in-plane spin accumulation. Importantly,
accumulation of the spin component parallel to $\bm m$ never occurs.
An externally imposed strain~\cite{Madhavan1,Madhavan2} breaks rotational symmetry, and is expected to induce the $u_3$ and $u_4$ terms, allowing control of the
direction of the spin polarization. The richness of the exhibited behavior allows targeted control of the interface properties in SE-TI systems.

{\it Discussion and Concluding Remarks}
We have shown that the properties of electronic states at SE-TI interfaces are determined by the symmetries of the interface, and not simply by topological considerations. Only under special conditions are these states described by the simplest Dirac Hamiltonian, possessing perfect helicity. In general SE-TI interfaces are characterized by an anisotropic energy dispersion, with the electron spin locked at an interface-dependent angle relative to the momentum, see Eq.~\eqref{hvsPP}. This implies dramatic control of the spin polarization and the gap in the spectrum. Our results suggest that some of the conclusions about the properties of topological heterostructures~\cite{FU-Kane,Balents,Xiao-Liang,Rossi} may need to be revisited.

First principles
calculations
are required to precisely determine the values of the $u_i$ for
specific interfaces.
In conventional semiconductor heterostructures
the relaxation of atomic positions leads to a lowering of the crystal symmetry to $C_{2v}$ at the interface irrespective of the bulk symmetry \cite{Zunger,Voisin2}. Our analysis is consistent with this picture.
Dependence of the interface potential on the in-plane coordinates, $x$ and $y$ (e.g. due to buckling) gives higher order terms in
$p_{x,y}$,
relevant only at higher energies (similar to trigonal lattice potentials~\cite{Fu2009}).

In summary, our analysis yields the most general low-energy (linear in momentum) Hamiltonian, Eq.~(\ref{hvsPP}),  that describes topological interface states, and lays the foundation for further work.
Our results suggest the tantalizing possibility of tuning the effective low energy Hamiltonian by experimental design of the  interface properties, enriching and enhancing the range of applications of topological insulators.

{\it Acknowledgements.}
This research was supported by NSF via grants DMR-1410741 and DMR-1151717.

\bibliographystyle{apsrev4-1}

\widetext
\newpage
\begin{center}
\textbf{\large Supplementary Material: Interface symmetry and spin control in topological insulator-semiconductor heterostructures}
\end{center}
\setcounter{equation}{0}
\setcounter{figure}{0}
\setcounter{table}{0}
\setcounter{page}{1}
\makeatletter
\renewcommand{\theequation}{S\arabic{equation}}
\renewcommand{\thefigure}{S\arabic{figure}}
\renewcommand{\bibnumfmt}[1]{[S#1]}
\renewcommand{\citenumfont}[1]{S#1}

\section{Particle-hole symmetry and the spins: the vector $\mathbf{m}$}

Particle-hole symmetry is present if there is an operator anticommuting with the Hamiltonian. Consider an operator $M=\bm m\cdot\bm\sigma$, such that for any momentum $\bm p=(p_x, p_y, 0)$ we have $\{H'_{I},M\}=0$ with
\begin{equation}
  H'_{I}=H_{I}-\Delta\sigma_{0}=\sum_{m=x,y,z}\sum_{n=x,y} c_{mn}\sigma_m p_n\equiv\bm c_{\bm p}\cdot\bm\sigma,
\end{equation}
where $\bm c_{\bm p, m}=\sum_n c_{mn} p_n$. Anticommutation requires that $\bm m\cdot \bm c_{\bm p}=0$ for any $\bm p$, which is satisfied for the unit vector
\begin{equation}\label{mvec}
\widehat{\bm m}\equiv\frac{ \bm m}{|\bm m|}= -\frac{\big(c_{yy}c_{zx} - c_{yx} c_{zy}\big) \hat{x} + \big(c_{xx}c_{zy} - c_{xy} c_{zx}\big) \hat{y}
+ \big(c_{xy}c_{yx} - c_{xx} c_{yy}\big) \hat{z}}{\sqrt{\big(c_{yy}c_{zx} - c_{yx} c_{zy}\big)^2+\big(c_{xx}c_{zy} - c_{xy} c_{zx}\big)^2+\big(c_{xy}c_{yx} - c_{xx} c_{yy}\big)^2}}\;.
\end{equation}
Since the spin eigenstates of $H_I^\prime$ and $H_I$ have spins pointing along or opposite $\bm c_{\bm p}$, it follows that spins of all the eigenstates are in the plane normal to $\bm m$.



\section{Transformation of the Hamiltonian to the form suitable for symmetry analysis.}

Equating the coefficients of Eq. (2) and Eq.(3) of the main text for the choice $\bf e=(\sin\phi,\cos\phi, 0)$ and $\bar{\bf e}=\widehat{\bm z}\times {\bf e}=(\cos\phi, -\sin\phi, 0)$ gives
\begin{eqnarray}
  v_6&=&(c_{xy}+c_{yx})\sin 2\phi + (c_{yy}-c_{xx})\cos 2\phi\,,
  \\
  v_5&=& c_{zx}\cos\phi - c_{zy}\sin\phi\,,
  \\
  v_4&=&(c_{xy}+c_{yx})\cos 2\phi - (c_{yy}-c_{xx})\sin 2\phi\,,
  \\
  v_3&=& c_{zx}\sin\phi + c_{zy}\cos\phi\,,
  \\
  v_2&=& \frac{1}{2}\Bigl[(c_{yy}+c_{xx})-v_6\Bigr]\,,
  \\
  v_1&=&\frac{1}{2}\Bigl[(c_{xy}-c_{yx})-v_4\Bigr]\,.
\end{eqnarray}
When the vector $\bf e$ is selected by the interface potential, $\widehat U$, and the coefficients $c_{mn}$ of the effective surface hamiltonian are also determined by the same potential and the bulk parameters, there is a unique choice of the coefficients $v_i$, which follows from the solution of the boundary problem, described below.
Similarly, for a given $\widehat U$,  the vectors $\bm m$ and $\bf e$ are fixed relative to each other.

\section{Determining $v_{i}$ from the Microscopic Model. }

\subsection{Boundary value problem and numerical calculations.}

To obtain the coefficients $v_{i}$ for a given set of the parameters $u_{i}$ defining the interface potential, we solve the boundary value problem, Eq. (4) in the main text. We start by finding the eigenfunctions corresponding to the bulk states of the TI and the SE that decay away from the interface, which satisfy $H_{\rm SE/TI}\psi=E\psi$, with $H_{\rm SE/TI}$ given in Eq.(5) of the main text. Explicitly,
\begin{small}
\begin{equation}\label{diff}
  \left(
    \begin{array}{cccc}
      M_{\mu}+B_{1}\partial^{2}_{z}-B_{2}p^{2}&-A_{1}\partial_{z} & 0& iA_{2}p_{-} \\
      A_{1}\partial_{z} & -(M_{\nu\mu}+B_{1}\partial^{2}_{z}-B_{2}p^{2})& iA_{2}p_{-}  & 0 \\
      0 & -iA_{2}p_{+} &M_{\nu\mu}+B_{1}\partial^{2}_{z}-B_{2}p^{2}&-A_{1}\partial_{z} \\
       -iA_{2}p_{+} & 0& A_{1}\partial_{z}&  -(M_{\nu\mu}+B_{1}\partial^{2}_{z}-B_{2}p^{2}) \\
    \end{array}
  \right)\psi
  =E\psi\,
\end{equation}
\end{small}
where $\mu=(B,T)$ denotes bottom/top of the heterostructure, $M_{B}=M$, $M_{T}=-m$, $p_\pm=p_x\pm i p_y$, and $p^2=p_x^2+p_y^2$.
Since the Hamiltonian commutes with the helicity operator, $\hat{h}\psi_{t\nu\mu}=t\psi_{t\nu\mu}$, the wave functions, $\psi$,  can be chosen to be simultaneously the eigenfunctions of $\hat{h}$, and written as
\begin{equation}
\psi_{t\nu\mu}(x,y,z)=\left(
                                   \begin{array}{c}
                                     ia_{t\nu\mu} \\
                                     ib_{t\nu \mu} \\
                                     t a_{t\nu\mu}e^{i\theta_{p}} \\
                                    t b_{t\nu\mu}e^{i\theta_{p}}  \\
                                   \end{array}\right)e^{i\bm p\cdot \bf r}e^{\lambda_{\nu\mu}z}\;,
\end{equation}
such that $\hat{h}\psi_{t\nu\mu}=t\psi_{t\nu\mu}$, 
with $t=\pm$. Here  $\bm p$  and $\bf r$ are in the $x$-$y$ plane, $\theta_{p}=\tan^{-1}(p_{y}/p_{x})$, and
\begin{subequations} \label{as}
\begin{eqnarray}
  a_{t\nu\mu} &=& A_{1}\lambda_{\nu\mu}-t A_{2}p \;,\\
  b_{t\nu\mu} &=& M_{\mu}+B_{1}\lambda^{2}_{\nu\mu}-B_{2}p^{2}-E\;.
\end{eqnarray}
\end{subequations}

In order to satisfy Eq.~\ref{diff}, the decay exponents of the interface state, $\lambda_{\nu\mu} (E,\bm p)$, must be the roots of Eq.(13) in the Methods section,
 \begin{equation}\label{lambda}
E^2-\mathcal{M}_{+\mu}\mathcal{M}_{-\mu}-A^{2}_{2}p^{2}=0\;,
\end{equation}
where $\mathcal{M}_{\pm\mu}=M_{\mu}+B_{1}\lambda^{2}_{\mu}-B_{2}p^{2}\pm A_{1}\lambda_{\mu}$. This gives
\begin{small}
\begin{subequations}\label{lambdas}
\begin{eqnarray}
  \lambda_{\nu B}&=& \frac{\sqrt{A^{2}_{1}-2B_{1}M+2B_{1}B_{2}p^{2}+\nu\sqrt{A^{4}_{1}-4A^{2}_{1}B_{1}(M-B_{2}p^{2})+4B^{2}_{1}(E^{2}-A^{2}_{2}p^{2})}} }{\sqrt{2}|B_{1}|}\;, \\
\lambda_{\nu T}&=& \frac{-\sqrt{A^{2}_{1}+2B_{1}m+2B_{1}B_{2}p^{2}+\nu\sqrt{A^{4}_{1}+4A^{2}_{1}B_{1}(m+B_{2}p^{2})+4B^{2}_{1}(E^{2}-A^{2}_{2}p^{2})}}}{\sqrt{2}|B_{1}|}\;,
\end{eqnarray}
\end{subequations}
\end{small}

The wave function, Eq.(12) of the Methods section, is the solution of the full interface problem if it satisfies the boundary conditions at $z=0$
\begin{subequations}\label{BC}
\begin{eqnarray}
& \sum_{t\nu}{C_{t\nu B}\psi_{t\nu B}(x,y,0)}-\sum_{t\nu}{C_{t\nu T}\psi_{t\nu T}(x,y,0)}= 0\;,& \\
& B_{1}\left( \sum_{t\nu }{C_{t\nu  B}\lambda_{\nu B}\psi_{t\nu  B}(x,y,0)}-\sum_{t\nu }{C_{t\nu  T}\lambda_{\nu T}\psi_{t\nu  T}(x,y,0)}\right)-\tau_{z}\hat{U}\psi(x,y,0) = 0&\;.
\end{eqnarray}
\end{subequations}
This boundary value problem is written in a matrix form as
\begin{equation}\label{usbvp}
\mathcal{B}x=\left[\mathcal{B}_{0}-(\mathcal{B}_{\tau_{0}}+\mathcal{B}_{\tau_{z}}+\mathcal{B}_{\tau_{x}}+\mathcal{B}_{5}+\mathcal{B}_{q})\right]x=0\;,
 \end{equation}
 where
  \begin{tiny}
\begin{equation}\label{m08}
\mathcal{B}_{0}=\left(
        \begin{array}{cccccccc}
         a_{++B} &  a_{+-B} &  a_{-+B} &  a_{--B} & -a_{++T} &  -a_{+-T} &  -a_{-+T} &  -a_{--T}  \\
          b_{++B} & b_{+-B} &  b_{-+B} &  b_{--B} & -b_{++T} &  -b_{+-T} &  -b_{-+T} &  -b_{--T}  \\
          a_{++B} &  a_{+-B} &  -a_{-+B} & - a_{--B} & -a_{++T} &  -a_{+-T} &  a_{-+T} &  a_{--T}\\
          b_{++B} &  b_{+-B} &  -b_{-+B} & - b_{--B} & -b_{++T} &  -b_{+-T} &  b_{-+T} &  b_{--T} \\
         B_{1}\lambda_{+B}a_{++B} &  B_{1}\lambda_{-B}a_{+-B} &  B_{1}\lambda_{+B}a_{-+B} &  B_{1}\lambda_{-B}a_{--B} & -B_{1}\lambda_{+T}a_{++T} &  -B_{1}\lambda_{-T}a_{+-T} &  -B_{1}\lambda_{+T}a_{-+T} &  -B_{1}\lambda_{-T}a_{--T}  \\
         B_{1}\lambda_{+B}b_{++B} &  B_{1}\lambda_{-B}b_{+-B} &  B_{1}\lambda_{+B}b_{-+B} &  B_{1}\lambda_{-B}b_{--B} & -B_{1}\lambda_{+T}b_{++T} &  -B_{1}\lambda_{-T}b_{+-T} &  -B_{1}\lambda_{+T}b_{-+T} &  -B_{1}\lambda_{-T}b_{--T}  \\
         B_{1}\lambda_{+B}a_{++B} &  B_{1}\lambda_{-B}a_{+-B} &  -B_{1}\lambda_{+B}a_{-+B} &  -B_{1}\lambda_{-B}a_{--B} & -B_{1}\lambda_{+T}a_{++T} &  -B_{1}\lambda_{-T}a_{+-T} &  B_{1}\lambda_{+T}a_{-+T} &  B_{1}\lambda_{-T}a_{--T} \\
         B_{1}\lambda_{+B}b_{++B} &  B_{1}\lambda_{-B}b_{+-B} &  -B_{1}\lambda_{+B}b_{-+B} &  -B_{1}\lambda_{-B}b_{--B} & -B_{1}\lambda_{+T}b_{++T} &  -B_{1}\lambda_{-T}b_{+-T} &  B_{1}\lambda_{+T}b_{-+T} &  B_{1}\lambda_{-T}b_{--T} \\
        \end{array}
      \right)\;,
\end{equation}
\end{tiny}
\begin{tiny}
\begin{equation}\label{b80}
\mathcal{B}_{\tau_{0}}=\frac{u_{0}}{2}\left(
                  \begin{array}{cccccccc}
                    0 & 0 &0 & 0 &0 & 0 & 0 & 0 \\
                    0 & 0 & 0 & 0 & 0 & 0 & 0 & 0 \\
                     0 & 0 & 0 & 0 & 0 & 0 & 0 & 0 \\
                    0 & 0 & 0 & 0 & 0 & 0 & 0 & 0 \\
                    a_{++B}& a_{+-B} & a_{-+B} & a_{--B} & a_{++T} & a_{+-T}& a_{-+T} & a_{--T} \\
                    -b_{++B}& -b_{+-B} &-b_{-+B} &  -b_{--B} &-b_{++T} & -b_{+-T}& -b_{-+T} & -b_{--T} \\
                    a_{++B} & a_{+-B}& -a_{-+B} & -a_{--B} & a_{++T} &a_{+-T} & -a_{-+T} & -a_{--T} \\
                    -b_{++B} & -b_{+-B} & b_{-+B} & b_{--B} & -b_{++T} & -b_{+-T} & b_{-+T} & b_{--T} \\
                  \end{array}
                \right)\;,
\end{equation}
\begin{equation}\label{b8z}
\mathcal{B}_{\tau_{z}}=\frac{u_{1}}{2}\left(
                  \begin{array}{cccccccc}
                    0 & 0 &0 & 0 &0 & 0 & 0 & 0 \\
                    0 & 0 & 0 & 0 & 0 & 0 & 0 & 0 \\
                     0 & 0 & 0 & 0 & 0 & 0 & 0 & 0 \\
                    0 & 0 & 0 & 0 & 0 & 0 & 0 & 0 \\
                    a_{++B}& a_{+-B} & a_{-+B}  & a_{--B} & a_{++T} & a_{+-T}& a_{-+T} & a_{--T} \\
                    b_{++B}& b_{+-B} &b_{-+B} & b_{--B} &b_{++T} & b_{+-T}& b_{-+T} & b_{--T} \\
                    a_{++B} & a_{+-B}& -a_{-+B} & -a_{--B} &  a_{++T} & a_{+-T} &- a_{-+T} & -a_{--T} \\
                   b_{++B} & b_{+-B}& -b_{-+B} & -b_{--B} & b_{++T} &  b_{+-T} & -b_{-+T} & -b_{--T} \\
                  \end{array}
                \right)\;,
\end{equation}
\begin{equation}\label{b8x}
\mathcal{B}_{\tau_{x}}=\frac{u_{2}}{2}\left(
                  \begin{array}{cccccccc}
                    0 & 0 &0 & 0 &0 & 0 & 0 & 0 \\
                    0 & 0 & 0 & 0 & 0 & 0 & 0 & 0 \\
                     0 & 0 & 0 & 0 & 0 & 0 & 0 & 0 \\
                    0 & 0 & 0 & 0 & 0 & 0 & 0 & 0 \\
                   b_{++B}& b_{+-B} & b_{-+B} & b_{--B} & b_{++T} & b_{+-T}& b_{-+T} & b_{--T} \\
                    -a_{++B}&- a_{+-B} & -a_{-+B} & -a_{--B} &-a_{++T} & -a_{+-T}&-a_{-+T} & - a_{--T} \\
                    b_{++B} &  b_{+-B}& -b_{-+B} & -b_{--B} & b_{++T} & b_{+-T} &-b_{-+T} & -b_{--T} \\
                    -a_{++B} & - a_{+-B} & a_{-+B} & a_{--B} & -a_{++T} & -a_{+-T} & a_{-+T} & a_{--T} \\
                  \end{array}
                \right)\;.
\end{equation}
\end{tiny}
\begin{tiny}
\begin{equation}\label{sigz}
\mathcal{B}_{5}=\frac{iu_{5}}{2}\left(
                 \begin{array}{cccccccc}
    0 & 0 & 0 & 0& 0 &0 & 0 & 0 \\
     0 & 0 & 0 & 0& 0 &0 & 0 & 0 \\
    0 & 0 & 0 & 0& 0 &0 & 0 & 0\\
     0 & 0 & 0 & 0& 0 &0 & 0 & 0 \\
                   -  b_{++B} & -b_{+-B} &  -b_{-+B} & -b_{--B} &  - b_{++T} & -b_{+-T} &  -b_{-+T} & -b_{--T} \\
                   -a_{++B}& -a_{+-B}&  -a_{-+B} & -a_{--B} &  -a_{++T} & -a_{+-T} &  -a_{-+T} & -a_{--T}\\
                    b_{++B} & b_{+-B} &  -b_{-+B}& -b_{--B} &  -b_{++T} & -b_{+-T} &  b_{-+T} &b_{--T} \\
                   a_{++B}& a_{+-B} &  -a_{-+B} & -a_{--B} &  -a_{++T} & -a_{+-T} &  a_{-+T} & a_{--T}\\
                 \end{array}
               \right)\;,
\end{equation}
\end{tiny}
\begin{tiny}
\begin{equation}\label{mq}
\mathcal{B}_{q}=\frac{iq}{2}\left(
  \begin{array}{cccccccc}
    0 & 0 & 0 & 0& 0 &0 & 0 & 0 \\
     0 & 0 & 0 & 0& 0 &0 & 0 & 0 \\
    0 & 0 & 0 & 0& 0 &0 & 0 & 0\\
     0 & 0 & 0 & 0& 0 &0 & 0 & 0 \\
    b_{++B}e^{i(\theta_{p}+\phi)} & b_{+-B}e^{i(\theta_{p}+\phi)} & -b_{-+B}e^{i(\theta_{p}+\phi)} & -b_{--B}e^{i(\theta_{p}+\phi)} & b_{++T}e^{i(\theta_{p}+\phi)} & b_{+-B}e^{i(\theta_{p}+\phi)} & -b_{-+B}e^{i(\theta_{p}+\phi)} & -b_{--B}e^{i(\theta_{p}+\phi)} \\
    a_{++B}e^{i(\theta_{p}+\phi)} & a_{+-B}e^{i(\theta_{p}+\phi)} & -a_{-+B}e^{i(\theta_{p}+\phi)} & -a_{--B}e^{i(\theta_{p}+\phi)} & a_{++T}e^{i(\theta_{p}+\phi)} & a_{+-B}e^{i(\theta_{p}+\phi)} & -a_{-+B}e^{i(\theta_{p}+\phi)} & -a_{--B}e^{i(\theta_{p}+\phi)}\\
 b_{++B}e^{-i(\theta_{p}+\phi)} & b_{+-B}e^{-i(\theta_{p}+\phi)} & b_{-+B}e^{-i(\theta_{p}+\phi)} & b_{--B}e^{-i(\theta_{p}+\phi)} & b_{++T}e^{-i(\theta_{p}+\phi)} & b_{+-B}e^{-i(\theta_{p}+\phi)} & b_{-+B}e^{-i(\theta_{p}+\phi)} & b_{--B}e^{-i(\theta_{p}+\phi)} \\
    a_{++B}e^{-i(\theta_{p}+\phi)} & a_{+-B}e^{-i(\theta_{p}+\phi)} &a_{-+B}e^{-i(\theta_{p}+\phi)} & a_{--B}e^{-i(\theta_{p}+\phi)} & a_{++T}e^{-i(\theta_{p}+\phi)} & a_{+-B}e^{-i(\theta_{p}+\phi)} & a_{-+B}e^{-i(\theta_{p}+\phi)} & a_{--B}e^{-i(\theta_{p}+\phi)} \\
  \end{array}
\right)\;,
\end{equation}
\end{tiny}
\begin{tiny}
with
\begin{equation}\label{xnon}
  x=\left(
      \begin{array}{c}
        C_{++ B} \\
         C_{+- B} \\
         C_{-+ B} \\
         C_{-- B} \\
        C_{++ T} \\
         C_{+- T} \\
         C_{-+ T} \\
         C_{-- T} \\
      \end{array}
    \right)\;.
\end{equation}
\end{tiny}
In order to have non-trivial solutions for $C_{t\nu \mu }$, we need det${\cal B}=0$
To solve this problem numerically find the eigenvalues $E$ at each in-plane momentum ${\bm p}$ from this condition, and determine the corresponding $C_{t\nu \mu }$ coefficients from the eigenvectors.

\subsection{Analytical approach and derivation of the surface Hamiltonian}

We now make a small energy and momentum expansion that allows us to find the analytical forms of the energy dispersion and the eigenvectors, which, in turn, gives  the analytical form of the interface Hamiltonian.
We first approximate the coefficients $a_{t\nu\mu}$ and $b_{t\nu\mu}$ (Eq.~\ref{as}) up to linear order in $E$ and $p$. Since the expansion of $\lambda_{\nu\mu}$, Eq.~(\ref{lambdas}), starts with the quadratic terms,
we replace these lengths by their values at $E=p=0$,
\begin{subequations}\label{aproxlam}
\begin{eqnarray}
  \lambda_{\nu B} &=&\frac{A_{1}+\nu\sqrt{A^{2}_{1}-4B_{1}M}}{2B_{1}}\;, \\
  \lambda_{\nu T} &=&-\frac{\nu A_{1}+\sqrt{A^{2}_{1}+4B_{1}m} }{2B_{1}}\;.
\end{eqnarray}
\end{subequations}
Consistently with this (and our expectation of linear dispersion), we omit the $B_2p^2$ term in the expressions for the coefficients $  a_{t \nu \mu}$ and $b_{t \nu \mu}$ , which gives
\begin{subequations}\label{approxab}
\begin{eqnarray}
  a_{t \nu B} &=& \frac{A^{2}_{1}+\nu A_{1}\sqrt{A^{2}_{1}-4B_{1}M}}{2B_{1}}-t A_{2}p\;, \\
  a_{t \nu T} &=&- \frac{\nu A^{2}_{1}+ A_{1}\sqrt{A^{2}_{1}+4B_{1}m} }{2B_{1}}-t A_{2}p\;,  \\
  b_{t \nu B} &=& M+\left(\frac{A_{1}+\nu\sqrt{A^{2}_{1}-4B_{1}M}}{2\sqrt{B_{1}}} \right)^2-E \\
   b_{t \nu T}  &=& -m+\left(\frac{\nu A_{1}+\sqrt{A^{2}_{1}+4B_{1}m} }{2\sqrt{B_{1}}} \right)^2-E\;.
\end{eqnarray}
\end{subequations}

Substituting the coefficients in Eq.~(\ref{approxab}),
we expand $\mbox{det} \mathcal{B}$ up to second order in both $E$ and $p$, and find the electronic dispersion, $E(\bm p)$ as well as coefficients $C_{t\nu\mu}$. This determines the eigenfunctions $\psi_{B}(x,y,z)$ and $\psi_{T}(x,y,z)$ in Eq.(13) of the main text.


We find the interface state by setting  $z=0$ in the wave functions, determining
 $\psi(x,y)=\psi_{B}(x,y,0)=\psi_{T}(x,y,0)$, and using it to form a 4$\times$4 Hamiltonian matrix.
That matrix has rank 2. Therefore, in order to find the effective interface Hamiltonian in the spin basis, we first apply the unitary transformation $R=e^{i\tau_{y}\sigma_{0}\vartheta}$ to the $4$-component interface spinor $\psi(x,y)$. The presence of $\tau_y$ in this rotation means that the wave functions of different parity are mixed (inversion symmetry is broken), but the spin remains a good quantum number (time-reversal is preserved).  This unitary transformation eliminates two of the spinor components, such that
\begin{equation}\label{ru0}
R\psi_{t}(x,y)=\tilde\psi_{t}(x,y)=\left(
                  \begin{array}{c}
                    \tilde\psi_{\uparrow} \\
                    0\\
                   \tilde\psi_{\downarrow} \\
                   0
                  \end{array}
                \right)e^{i {\bm p}\cdot {\bf r}}\;,
\end{equation}
where $\psi_{t=\pm}$ are the eigenvectors corresponding to $E+\Delta$ and $-E+\Delta$. Then in the space defined by the operators
\begin{equation}\label{newfieldop0}
\left(
             \begin{array}{c}
               \bar{\psi}_{\uparrow} \\
               \bar{\psi}_{\downarrow}\\
             \end{array}\right)=\left(
             \begin{array}{c}
               \hat{\psi}_{+\uparrow}\cos(\vartheta)+\hat{\psi}_{-\uparrow}\sin(\vartheta) \\
              \hat{ \psi}_{+\downarrow}\cos(\vartheta)+\hat{\psi}_{-\downarrow}\sin(\vartheta) \\
             \end{array}
           \right)
\;.
\end{equation}
the Hamiltonian for the interface takes the form of Eq.(3)
\begin{equation}\label{fistsurface0}
H_{I}=\Delta\sigma_0+  v_1 ( \bm{\sigma} \times \mathbf{p})_z + v_2 \bm{\sigma} \cdot \mathbf{p} + v_3 \sigma_z \mathbf{p}\cdot \mathbf{e}+ v_4 ( \bm{\sigma} \times \mathbf{e})_z \mathbf{p}\cdot\mathbf{e}+ v_5  \sigma_z ( \mathbf{p}\times\mathbf{e})_z+ v_6 (\bm{\sigma}\cdot \mathbf{e}) ( \mathbf{p}\cdot \mathbf{e}) \;.
\end{equation}

We now consider several representative cases.

\subsubsection{\bf ${\cal A}_1$ irrep: $\hat{U}=\sigma_{0}\otimes(u_{0}\tau_{0}+u_{1}\tau_{z}+ u_{2}\tau_{x})$.}

In this case the dispersion relation up to linear order $p$ is given by
\begin{equation}\label{disp1}
  E=\pm\mathcal{F}A_{2}p+\Delta\;,
\end{equation}
where $A_2$ is the band parameter in Eq.(5), and
\begin{small}
\begin{subequations}\label{shiftandslope}
\begin{eqnarray}
\Delta&=&-\frac{4 A_{1} (u_{0} + u_{2}) m M}{\beta_{0}}\;, \\
\mathcal{F}&=&-\frac{\beta_{1}}{\beta_{0}}\;, \\
  \beta_{0}&=&  2A_{1}(u_{0}+u_{2})(M-m)+2B_{1}\lambda_{+T}(A_{1}(m+M)-2u_{1}m)-2m(u^{2}_{1}+u^{2}_{2}-u^{2}_{0}+B_{1}(m+M))\\
  \beta_{1}&=&-2(u_{0}+u_{2})(2B_{1}M\lambda_{+T}+A_{1}(m+M))-2B_{1}\lambda_{+T}(A_{1}(m+M)-2u_{1}m)+2m(u^{2}_{1}+u^{2}_{2}-u^{2}_{0}+B_{1}(m+M))\;.
\end{eqnarray}
\end{subequations}
\end{small}
The interface state spinor is
\begin{equation}\label{s1}
\psi_{t}(x,y,0)=\frac{1}{N}\left(
  \begin{array}{c}
  iA_{1}\lambda_{-T}\lambda_{+T}(u_{0}+u_{2}-u_{1}+B_{1}(\lambda_{+B}-\lambda_{+T})) \\
 i((u_{0}+u_{2})\lambda_{-T}(\Delta-A_{1}\lambda_{+T})-(\Delta+A_{1}\lambda_{-T})(u_{1}+B_{1}(\lambda_{+T}-\lambda_{+B})))\\
tA_{1}\lambda_{-T}\lambda_{+T}(u_{0}+u_{2}-u_{1}+B_{1}(\lambda_{+B}-\lambda_{+T}))e^{i\theta_{p}}\\
t((u_{0}+u_{2})\lambda_{-T}(\Delta-A_{1}\lambda_{+T})-(\Delta+A_{1}\lambda_{-T})(u_{1}+B_{1}(\lambda_{+T}-\lambda_{+B})))e^{i\theta_{p}} \\
  \end{array}
\right)e^{i\bm p\cdot \bm r}
\end{equation}
where $t={\rm sgn}(E-\Delta)$, and $N$ is the normalization factor. Clearly only the first and the second components of the spinor are linearly independent, so that rotation by
\begin{equation} \vartheta=\tan^{-1}\left(\frac{((u_{0}+u_{2})\lambda_{-R}(\Delta-A_{1}\lambda_{+R})-(\Delta+A_{1}\lambda_{-R})(u_{1}+B_{1}(\lambda_{+R}-\lambda_{+L})))}{A_{1}\lambda_{-R}\lambda_{+R}(u_{0}+u_{2}-u_{1}+B_{1}(\lambda_{+L}-\lambda_{+R}))}\right)\;,
 \end{equation}
  brings the Hamiltonian to the desired form,
\begin{equation}\label{fistsurface}
H_I=\Delta\sigma_{0}+v_{1}(\bm{\sigma}\times \bm{p})_{\hat{z}}
\end{equation}
with $v_{1}=A_{2}\mathcal{F}$.

\subsubsection {\bf ${\cal A}_2$ irrep: $\hat{U}=u_{5}\sigma_{z}\otimes\tau_{y}$.}

In this case the energy shift vanishes, and the dispersion relation is given by
\begin{eqnarray}\label{u2disp}
E&=&\pm\mathcal{F}_{5}A_{2}p\;,
\\
\label{u5disp1}
\mathcal{F}_{5}&=&\frac{\sqrt{(\mathcal{G}+\mathcal{G}_{5})^2-\frac{4M\mathcal{G}\mathcal{G}_{5}}{(m+M)}}}{\mathcal{G}+\mathcal{G}_{5}}\;,
\\
  \mathcal{G} &=& 2(B_{1}\lambda_{+T})^{2}(m+M)\,, \\
  \mathcal{G}_{5} &=& 2m u_{5}^2\;.
\end{eqnarray}
The interface spinor,
\begin{equation}\label{su5}
\psi(x,y,0)=\frac{1}{\sqrt{2}} \left(
   \begin{array}{c}
   i \\
       i   \\
      {\rm sgn}(E) e^{i(\theta_{p}-\beta)}  \\
      {\rm sgn}(E)e^{i(\theta_{p}-\beta) }\\
   \end{array}
 \right)e^{i\bm{p}\cdot \bm{r}}\;,
\end{equation}
with
\begin{equation}\label{beta}
 \frac{\beta}{2} = \tan^{-1}\left(\frac{u_{5}(\lambda_{-B}+\lambda_{-T})}{(m+M)} \right)\;,
  \end{equation}
 obviously has only two independent components. Then rotation by $\vartheta=\pi/4$ gives
 the effective interface Hamiltonian
\begin{equation}\label{ham5}
H_I=v_{1}(\bm{\sigma}\times \bm{p})_{\hat{z}}+v_{2}\bm{\sigma}\cdot \bm{p }\;,
 \end{equation}
with
\begin{subequations}
\begin{eqnarray}
  v_{1} &=& A_{2}\mathcal{F}_{5}\cos(\beta) \\
  v_{2}&=&-A_{2}\mathcal{F}_{5}\sin(\beta)\;.
\end{eqnarray}
\end{subequations}

 \subsubsection{\bf ${\cal B}_2$ irrep: $\hat{U}=(u_{3}\sigma_{x}+u_{4}\sigma_{y})\otimes\tau_{y} \equiv q\sigma_{y}e^{-i \sigma_{z}\phi}\otimes \tau_{y}$.}

 In this final case the eigenvalues take the form
\begin{equation}\label{dispelip}
E=\pm A_{2}p\mathcal{F}_{\theta_{p}}\;,
\end{equation}
where
\begin{subequations}\label{elip}
\begin{eqnarray}
\mathcal{F}_{\theta_{p}}&=&\frac{\sqrt{\mathcal{G}^{2}_{1}+\mathcal{G}_{q}^{2}+\mathcal{G}_{1}\mathcal{G}_{q}+\mathcal{G}_{1}\mathcal{G}_{q}\cos(2(\theta_{p}+\phi))}}{\mathcal{G}_{1}+\mathcal{G}_{q}}\;,\\
  \mathcal{G}_{1} &=& 2 \mathcal{G} \\
  \mathcal{G}_{q} &=& 2\sqrt{2}mq^2\;.
\end{eqnarray}
\end{subequations}

The approximate interface eigenstates,
\begin{equation}\label{sq}
  \psi(x,y,0)=\frac{1}{N}\left(
              \begin{array}{c}
              i({\rm sgn}(E)(\mathcal{G}_{1}+\mathcal{G}_{q})\mathcal{F}_{\theta_{p}} -\sqrt{2\mathcal{G}_{1}\mathcal{G}_{q}}\sin(\theta_{p}+\phi))e^{i(\theta_{p}+2\phi)}\\
               i({\rm sgn}(E)(\mathcal{G}_{1}+\mathcal{G}_{q})\mathcal{F}_{\theta_{p}} -\sqrt{2\mathcal{G}_{1}\mathcal{G}_{q}}\sin(\theta_{p}+\phi))e^{i(\theta_{p}+2\phi)}\\
              \mathcal{G}_{q}+\mathcal{G}_{1}e^{2i(\theta_{p}+\phi)}\\
                 \mathcal{G}_{q}+\mathcal{G}_{1}e^{2i(\theta_{p}+\phi)}
              \end{array}
            \right)e^{i \bm{p}\cdot \bm{r}}
\end{equation}
with $N$ again the normalization constant, once again clearly have only two independent components. Choosing $\vartheta=\pi/4$ as before  we find
\begin{equation}\label{hamqphi}
H_I=v_{1}(\bm{\sigma}\times \bm{p})_{\hat{z}}+(v_{3}\sigma_{z}+v_{4}(\bm{\sigma}\times \bm{e}))\bm{p}\cdot\bm{e}\;,
\end{equation}
with
\begin{eqnarray}
  v_{1} &=& \frac{A_{2} \mathcal{G}_{1}}{\mathcal{G}_{1}+\mathcal{G}_{q}}\;, \\
  v_{3} &=&  -\frac{A_{2}\sqrt{2\mathcal{G}_{1}\mathcal{G}_{q}}}{\mathcal{G}_{1}+\mathcal{G}_{q}}\;, \\
  v_{4}&=&- \frac{A_{2}\mathcal{G}_{q}}{\mathcal{G}_{1}+\mathcal{G}_{q}}\;.
\end{eqnarray}

\section{Effects of In-Plane Magnetic Fields}


The in-plane magnetic field, ${\bf B}={\bf B}(\cos\theta,\sin\theta, 0)$ only couples to the interface states via the Zeeman coupling, $g\mu_{B}{\bf B}\cdot\bm\sigma\equiv \bm h\cdot\bm \sigma$, where $\mu_B$ is the Bohr magneton, and $g$ is the gyromagnetic ratio. Adding this term to the Hamiltonian, Eq.(2) to obtain
\begin{equation}\label{interfacehamlt}
H_{I}({\bf B})=(c_{xx}\sigma_{x}+c_{yx}\sigma_{y}+c_{zx}\sigma_{z})p_{x}+(c_{xy}\sigma_{x}+c_{yy}\sigma_{y}+c_{zy}\sigma_{z})p_{y}+\Delta \sigma_{0}+\sigma_{x}h_{x}+\sigma_{y}h_{y}\;,
\end{equation}
and finding the eigenvalues, we determine the spectral gap
\begin{equation}\label{gap}
\delta({\bf B})=\frac{\left|c_{zx}(c_{yy}h_{x}-c_{xy}h_{y})+c_{zy}(c_{xx}h_{y}-c_{yx}h_{x})\right|}
{\sqrt{\big(c_{yy}c_{zx} - c_{yx} c_{zy}\big)^2+\big(c_{xx}c_{zy} - c_{xy} c_{zx}\big)^2+\big(c_{xy}c_{yx} - c_{xx} c_{yy}\big)^2}}.
\end{equation}
Comparison with Eq.~\eqref{mvec} shows that the gap can be rewritten in the form
\begin{equation}
  \delta({\bf B})=2 |\bm h\cdot\widehat{\bm m}|\,.
\end{equation}
For $\bm m\| \widehat{\bm z}$ (if both $c_{zx}$ and $c_{zy}$ vanish), the in-plane field does not open a gap. However, if the spins are not locked into the plane of the interface, when $\bm m$ acquires a component along the $z$-axis, the spectral gap depends on the relative orientation of the field and the in-plane projection $\bm m_\|=(m_x, m_y, 0)$. Maximal value of $\delta({\bf B})$ is reached for $\bf B$ along or opposite to $\bm B_\|$, while the gap vanishes identically for the field applied normal to $\bm m_\|$. 

For the specific case of only $u_{3,4}\neq 0$, $\bm m\cdot\bf e=0$, and therefore the gap vanishes for $\bf B$ along the direction of ${\bf e}$. In this case,
\begin{equation}\label{gapu3u4}
\delta({\bf B})=4h\sqrt{\frac{\mathcal{G}\mathcal{G}_{q}}{4\mathcal{G}^2+\mathcal{G}^{2}_{q}}}|\cos(\theta+\phi)|\;,
\end{equation}
where ${\bf e}=(\sin\phi, \cos\phi, 0)$.


\begin{thebibliography}{10}
\bibitem{v4} Such a term would immediately appear at any surface that does not preserve the full rotations, such as $(111)$ or $(100)$.
\end{thebibliography}
\end{document}